\documentclass[useAMS,usenatbib]{mn2e}

\title[Spatial and velocity coincidence of OH masers]
{Spatial and velocity coincidence of 4765- and 1720-MHz OH masers 
in two star-forming regions Cep A and W75N} 
\author[A. Niezurawska et al.]
       {A.~Niezurawska,$^1$ M.~Szymczak,$^1$
        R.J.~Cohen$^2$ and A.M.S.~Richards$^2$ \\
       $^1$Toru\'n Centre for Astronomy, Nicolaus Copernicus 
          University, Gagarina 11, 87-100 Toru\'n, Poland \\
       $^2$Jodrell Bank Observatory, University of Manchester,
          Macclesfield, Cheshire SK11 9DL, UK}

\date{Released 2003 Xxxxx XX}

\pagerange{\pageref{firstpage}--\pageref{lastpage}} \pubyear{2004}

\def\LaTeX{L\kern-.36em\raise.3ex\hbox{a}\kern-.15em
    T\kern-.1667em\lower.7ex\hbox{E}\kern-.125emX}

\usepackage{graphicx}
\begin{document}

\label{firstpage}

\maketitle

\begin{abstract}
We present the first maps of 4765-MHz OH masers in two star$-$forming
regions Cepheus\,A and W75N, made with  Multi-Element Radio Linked
Interferometer Network.  
In Cep\,A the emission has an arc$-$like structure of size 40\,mas 
with a clear velocity gradient, and lies at the edge of H\,{\small II} region 3b.
Over a period of 8 weeks the maser diminished in intensity by a factor of 7
but its structure remained stable. 
This structure coincides with a newly mapped 1720-MHz maser in Cep\,A within the 
positional error, and matches it in velocity.  
No emission of any other ground state or 4.7-GHz excited OH
transitions was detected in the 4765/1720-MHz region.
The 1720-MHz line exhibits Zeeman splitting that corresponds to a
magnetic field strength of $-$17.3\,mG.  
In W75N the excited 4765-MHz OH maser has a linear structure of size 45\,mas
with a well defined velocity gradient, and lies at the edge of H\,{\small
II} region VLA\,1. This structure coincides in position and
velocity with the 1720-MHz masers.  
We conclude that in both sources the 4765-MHz emission coexists
with 1720-MHz emission in the same volume of gas.
In such a case the physical conditions in these regions are
tightly constrained by the maser$-$pumping models.
\end{abstract}


\section{Introduction}
OH maser emission is an important probe of obscured star$-$forming 
regions. 
The maser lines provide high precision information on the location 
and velocity of the excited gas and on the physical conditions, notably 
the magnetic field (e.g.\,Cohen 1989 and references therein).  
If more than one OH maser transition is detected from the same compact 
region then it is possible to better constrain the other physical 
conditions such as gas temperature and density, the dust temperature, 
and so on (e.g.\,Gray et al. 2001 and references therein).  
However the detection of maser transitions of rotationally excited OH 
is not very efficient, with current equipment.  
Among three allowed transitions of the excited $^2\Pi_{1/2}$, $J$=1/2 state
of OH only the 4765-MHz ($F$=1$-$0) line is usually seen, but 
even this is very
rare (\citealt*{cohen91}; \citealt*{cohen95}; \citealt*{dodson02}).
A typical profile of a 4765-MHz source 
contains a single narrow unpolarized feature of 
intensity of a few hundred mJy which is usually variable 
on timescales that can be as short as a few weeks 
 (\citealt*{smits98}; \citealt{smits03}).  In fact several sources have 
been detected only during episodes of flaring. 
The two well known star-forming regions 
Cep\,A and W75N recently detected in the 4765-MHz line 
\citep*{szymczak00} appear to belong to this class of
objects that experience flaring.

\begin{table*}
\caption {Details of the MERLIN observations}   
\begin{tabular}{llclccrcc}
\hline\hline
       &       & &\multicolumn{2}{c}{\sc Phase$-$calibrator}&\multicolumn{2}{c}{\sc Synthesised Beam} &  & \\ 
       &       &            &      & Flux   &        &         &      & RMS \\
Source & Date  & Transition & Name & density&  HPBW  & P.A.    & Time &Noise \\
       &       &  (MHz)     &      & (mJy)  & (mas)  & (deg)   &  (h) &(mJy\,b$^{-1}$) \\
\hline
Cep A & 04Jan1999$^\star$ &  4765.562 & 2300$+$638 & 250  & 48$\times$38 & $-$14 & 10 & 2.5 \\
      & 27Jan1999$^\star$ &  4765.562 &            & 200  & 48$\times$40& $-$12 & 10 & 2 \\
      & 01Mar1999$^{\star\star}$ &  4660.242 &     & 230  &  & & 5 & \\
      &           &  4750.656 &     & 250  &  & & 5 & \\
      &           &  4765.562 &     & 250  &   49$\times$43 & $-$11 & 5 & 6 \\
      & 20May1999$^{\star\star}$ &   1720.530 & & 130 & 154$\times$107 & $-$24 & 7.5&3.5 \\
      & 16Jun1999$^{\star\star}$ &   1720.530 & & 150 & 143$\times$109 & $-$32 & 4 & 6 \\
W75N  & 14Apr2000$^{\star\star}$ & 4765.562 & 2005$+$403 & 2500 & 49$\times$45 & 59 & 7 & 3 \\ 
\hline\hline
\multicolumn{3}{l}{Antennas: $^\star$ Defford, Cambridge, Knockin, Darnhall,
Mark\,II} &
\multicolumn{6}{l}{$^{\star\star}$ Defford, Cambridge, Knockin, Darnhall,
Mark\,II, Tabley} \\
\end{tabular}
\end{table*}

Cep\,A is a part of the Cepheus OB3 association at an estimated
distance of 725\,pc \citep*{blaauw59}. The complex contains 
clusters of bright far infrared sources (e.g.\,\citealt{goetz98}),
compact radio continuum sources (e.g.\,\citealt*{hughes95};
\citealt{garay96}; \citealt{torrelles98}) and maser sources of several
species seen in many transitions (e.g.\,\citealt*{cohen84};
\citealt*{migenes92}; \citealt{torrelles98}; \citealt*{argon00}).
W75N lies in the molecular cloud complex DR21$-$W75 at a distance of 2
kpc \citep*{dickel78}.  Several radio continuum sources (e.g.\,\citealt{hunter94}; 
\citealt{torrelles97}) and maser
sources (\citealt{baart86}; \citealt{torrelles97};
\citealt*{hutawarakorn02}) are observed.  Shortly after the discovery of
the excited OH maser flares at 4765\,MHz in Cep\,A and W75N   
we undertook interferometric observations in 
order to determine the locations of the masers in
these very complex star$-$forming regions.

Single$-$dish studies have shown a correlation between 4765- and 1720-MHz
OH masers \citep{cohen95},  a connection that is 
supported by high angular resolution studies of few
selected sources (\citealt*{palmer84}; \citealt{baudry88}; \citealt{gray01}).
Recent observations at arcsecond resolution 
by \citet{dodson02} have suggested a 
stronger association between 6035- and 4765-MHz lines than
that between 1720- and 4765-MHz emission. 
Detailed modelling has shown that various combinations of OH transitions
can occur in restricted ranges of conditions in regions of shocked gas
(\citealt*{gray91}; \citealt*{gray92}; \citealt{pavlakis96a}; 
\citealt{pavlakis96b}; \citealt*{cragg02}). 
High angular resolution measurements and simultaneous 
identification of different OH maser transitions in the same volume 
of gas can be of a great value in estimating very precisely the local physical
conditions and distinguishing between different pumping models.  

In this paper we report on  multi-frequency observations of OH masers in 
the two objects. Our aims were to locate with high precision  
the excited and ground state OH masers and to determine their observational 
characteristics. The observations demonstrate for the first time the 
co-propagation of maser emission at 4765- and 1720\,MHz in both targets.    

\section{Observations and data reduction}
The observations of two star-forming regions Cep\,A and W75N were
carried out with 5 or 6 telescopes of MERLIN. 4765-MHz OH transition from 
Cep\,A was observed in three sessions between 1999 January and March and
from W75N in April 2000 (Table 1). Observations of the 1720-MHz OH line
towards Cep\,A were made at two epochs (Table 1). 
The other ground state transitions were also
observed and a comprehensive description of them will appear in a separate
paper. Each transition was observed simultaneously in left and right circular
polarization (LHC, RHC) and correlated to obtain all four Stokes
parameters.  The spectral bandwidth of 250-kHz was divided into
128 channels yielding a velocity resolution of
0.12\,km\,s$^{-1}$ at 4.7\,GHz and 0.34\,km\,s$^{-1}$ at
1720\,MHz. The velocity coverage was 16 and 45\,km\,s$^{-1}$ at
4.7\,GHz and 1720\,MHz respectively. The band centre was set to a
local standard of rest (LSR) velocity of $-$14\,km\,s$^{-1}$ for Cep A
and 10\,km\,s$^{-1}$ for W75N.

The phase$-$referencing technique was applied in all observations, with 
sources 2300+638 and 2005+403 used as phase calibrators for Cep A and
W75N, respectively (Table 1). The cycle times between Cep A and its 
phase reference were 8\,min+2\,min at the first 
and second epochs and 3\,min+2\,min at the third epoch for the 4.7\,GHz
transition and 6.5\,min+2\,min for the 1720\,MHz transition.  
For W75N and its phase reference source
the cycle time was 3.5\,min+2.5\,min. The reference sources were too weak 
to be observed in narrow$-$band mode, hence they were observed in wide$-$band
(16\,MHz) mode to obtain the required signal to noise ratio.

The data reduction was carried out with standard procedures
\citep{diamond03} using the local programmes at Jodrell Bank and the
Astronomical Image Processing System (AIPS).  3C286 was used as a
primary flux density calibrator \citep{baars77} and 3C84 as a point
source and bandpass calibrator. The 4.7-GHz data were imaged using a pixel
separation of 15-mas and a 40-mas circular Gaussian restoring beam;
for the 1720-MHz emission 40-mas pixels and a 120-mas beam were
used. In emission-free Stokes $I$ maps the rms noise levels ($\sigma$) were
typically few \,mJy\,beam$^{-1}$ (Table 1). 
Regions of 15$\arcsec\times$15$\arcsec$ and 20$\arcsec\times$20$\arcsec$
were searched for emission 4765- and 1720-MHz, respectively.
The positions and peak flux densities of maser spots brighter than 10$\sigma$
were determined by fitting 2D Gaussian components.

The absolute positional accuracy of maser components measured in the
paper was limited by a number of uncertainties: (i) the positional 
accuracy of the phase calibrators was about 12\,mas \citep{patnaik92}, (ii) 
the error arising from uncertainties in telescope positions was
about 10\,mas \citep{diamond03}, (iii) the angular separations between
targets and phase calibrators of 2$\degr$ (Cep A) and 2.8$\degr$
(W75N) gave errors of 13\,mas at 4765\,MHz, 30\,mas at 1720\,MHz
for Cep A and 10\,mas at 4765\,MHz for W75N, taking into account the worst
phase$-$rate for each observation. These combine to give total
absolute position errors for maser components in Cep A of 20 and
34\,mas at 4765- and 1720\,MHz respectively, and of 19\,mas for
4765-MHz masers in W75N. Where we are comparing observations using the
same phase-reference source and/or the same array, conditions (i)
and/or (ii) do not apply and only condition (iii) produces relevant 
1$\sigma$ systematic position errors. Each maser 
component also has a relative position error, given approximately by
beamsize/signal-to-noise-ratio, but this was negligible, being
$<0.5$-mas for a typical component.

\section{Results}
\subsection{Cep A}
\begin{figure*}
\resizebox{\hsize}{!}{\includegraphics{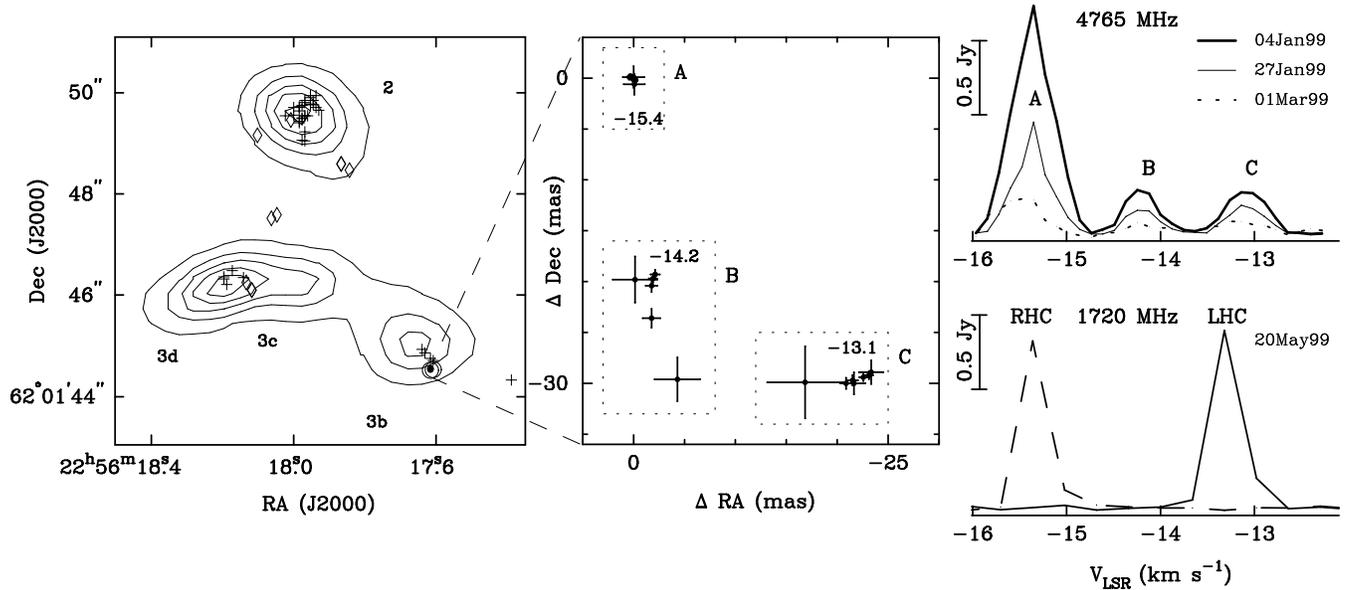}}
\caption{{\bf Left}: The 4.9-GHz VLA continuum map of the Cep\,A East
central region \citep{garay96} with contours at the levels of 10\%,
30\%, 50\%, 70\% and 90\% of 4.8\,mJy\,beam$^{-1}$.
The labels of continuum sources were taken from
\citet{hughes84}. The 1720-MHz and 4765-MHz OH maser components
reported in this paper are shown as circles and dots respectively.
The positions of the 22-GHz H$_2$O maser components (crosses) taken
from \citet{torrelles98} and those of the 1665-MHz OH maser components
(diamonds) \citep{argon00} are overlaid.  {\bf Middle:} Distribution
of 4765-MHz OH masers relative to the strongest component 
observed on 1999 January 4. Each of the three groups of components
are labelled by the central velocity V$_{\rm LSR}$ (\,km\,s$^{-1}$). The
bars indicate the relative errors in maser component positions.
{\bf Right:} The $I$ Stokes spectra of the 4765-MHz emission at three epochs
(top) and the 1720-MHz spectrum in left and right circular
polarization (bottom) from Cep A.}
\label{fig1} 
\end{figure*}

\begin{table*}
\caption {Components of the 4765-MHz OH maser observed towards Cep A at three 
epochs. The $\Delta$RA and $\Delta$Dec are given relative to the position
of the brightest component at the first epoch, 
i.e. RA(J2000) = 22$^{\rm h}$56$^{\rm m}$17\fs6176; 
Dec(J2000) = 62\degr01\arcmin44\farcs567.}
\begin{tabular}{l|rrc|rrc|rrc|c}
\hline\hline
Epoch        &\multicolumn{3}{|c|}{04Jan99} & \multicolumn{3}{c|}{27Jan99} & \multicolumn{3}{c}{01Mar99} \\
\hline
V$_c$        & $\Delta$RA & $\Delta$Dec & S$_{4765}$ & $\Delta$RA & $\Delta$Dec & S$_{4765}$ &
$\Delta$RA & $\Delta$Dec & S$_{4765}$ & Group  \\
(km\,s$^{-1}$) & (mas)        & (mas)         &(Jy\,b$^{-1}$)& (mas)  & (mas) &
 (Jy\,b$^{-1}$) & (mas)  & (mas) & (Jy\,b$^{-1}$) &\\
\hline
$-$15.84 & $-$0.1(1.0)  & $-$0.6(1.0)  & 0.105 & & & &       $-$0.3(3.1) & $-$3.3(3.3) & 0.133 & A \\
$-$15.72 &    0.4(0.3)  &    0.1(0.3)  & 0.042 & $-$0.1(0.5) & $-$0.4(0.5)& 0.127 & $-$1.8(2.3) & 0.1(2.2) & 0.166 & A \\
$-$15.60 &    0.1(0.1)  &    0.1(0.1)  & 0.841 &    0.0(0.2) & 0.0(0.2) & 0.305 & $-$2.1(1.9) & 0.4(1.9) & 0.227 & A \\
$-$15.47 &    0.2(0.1)  &    0.1(0.1)  & 1.225 &    0.4(0.2) & 0.0(0.2) & 0.446 & 0.1(1.7) & 0.3(1.6) & 0.235 & A \\
$-$15.35 &    0.0(0.1)  &    0.0(0.1)  & 1.554 &    0.0(0.1) & 0.0(0.1) & 0.744 & 0.9(1.7) & 0.0(1.6) &  0.231 & A \\
$-$15.23 &    0.3(0.1)  &    0.2(0.1)  & 1.109 & $-$0.1(0.2) & 0.0(0.2) & 0.387 & 1.1(6.3) & $-$1.6(7.0) & 0.073 & A \\  
$-$15.11 & $-$0.2(0.2)  & $-$0.2(0.2)  & 0.772 &    0.0(0.3) & 0.1(0.3) & 0.241 & & & & A \\
$-$14.98 &    0.1(0.3)  & $-$0.1(0.2)  & 0.378 & $-$0.1(0.6) & $-$0.6(0.6) & 0.095 & & & & A \\
$-$14.86 &    0.0(1.0)  &     0.1(0.3) & 0.096 & & & & & & & A \\
$-$14.49 & $-$0.1(2.2)  & $-$19.8(2.3) & 0.058 & & & & & & & B \\
$-$14.37 & $-$1.8(0.6)  & $-$20.4(0.6) & 0.216 & $-$3.7(0.6) & $-$15.9(0.6) & 0.111 & & & & B \\
$-$14.25 & $-$1.9(0.4)  & $-$19.7(0.4) & 0.304 & $-$1.1(0.4) & $-$18.4(0.4) & 0.151 & $-$4.9(5.9) & $-$23.2(6.9) & 0.074 & B \\
$-$14.12 & $-$2.1(0.5)  & $-$19.3(0.5) & 0.274 & $-$2.1(0.5) & $-$17.3(0.5) & 0.147 & $-$2.8(6.2) & $-$18.7(6.4) & 0.048 & B \\
$-$14.00 & $-$1.8(0.9)  & $-$23.6(0.9) & 0.136 & $-$6.3(1.0) & $-$18.2(1.0) & 0.071 & & & & B \\
$-$13.88 & $-$4.3(2.2)  & $-$29.6(2.2) & 0.055 & & & & & & & B \\
$-$13.51 & $-$16.8(3.7) & $-$29.9(3.5) & 0.033 & & & & & & & C  \\
$-$13.39 & $-$23.3(1.2) & $-$28.9(1.2) & 0.104 & $-$24.6(1.0) & $-$17.3(0.5) & 0.069 & $-$22.4(6.3) & $-$26.0(8.9) & 0.049 & C \\  
$-$13.26 & $-$21.5(0.5) & $-$29.7(0.5) & 0.236 & $-$18.3(0.5) & $-$28.4(0.5) & 0.124 & $-$22.9(6.1) & $-$28.7(6.5) & 0.072 & C \\
$-$13.14 & $-$22.6(0.4) & $-$29.4(0.4) & 0.290 & $-$19.5(0.4) & $-$29.9(0.3) & 0.197 & $-$18.5(4.7) & $-$27.8(5.0) & 0.076 & C \\  
$-$13.02 & $-$23.1(0.4) & $-$29.2(0.4) & 0.295 & $-$21.0(0.4) & $-$30.2(0.4) & 0.177 & $-$24.9(6.0) & $-$28.1(6.4) & 0.062 & C \\
$-$12.89 & $-$20.9(0.6) & $-$30.0(0.5) & 0.221 & $-$19.5(0.6) & $-$30.1(0.5) & 0.119 & & & & C \\
$-$12.77 & $-$21.6(1.0) & $-$30.0(1.0) & 0.106 & $-$22.4(1.3) & $-$30.0(1.2) & 0.056 & & & & C \\
\hline\hline
\end{tabular}                                
\end{table*} 

\begin{table}
\caption {Components of the 1720-MHz OH maser detected from Cep A on 1999
May 20. The $\Delta$RA and $\Delta$Dec are given similarly as in Table 2.}
\begin{tabular}{lrrcc}
\hline\hline
V$_c$        & $\Delta$RA & $\Delta$Dec & S$_{1720}$ & Polzn. \\
(km\,s$^{-1}$) & (mas)        & (mas)         &(Jy\,b$^{-1}$)& \\
\hline
 $-$15.36 &  $-$20.2 (0.4) & $-$50.0 (0.4) & 1.122 & RHC \\
 $-$15.02 &  $-$20.4 (4.0) & $-4$9.7 (4.0) & 0.121 & RHC \\ 
 $-$13.66 &  $-$45.7 (9.0) & $-$23.2 (9.0) & 0.059 & LHC \\
 $-$13.32 &  $-$36.5 (0.4) & $-$33.7 (0.4) & 1.050 & LHC \\
 $-$12.98 &  $-$37.8 (2.8) & $-$33.0 (3.0) & 0.210 & LHC \\
\hline\hline
\end{tabular}
\end{table}

The results for Cepheus A are summarized in Figure 1 and Tables 2 and 3.  
The 4765-MHz maser emission was found in the velocity range from $-$16
to $-$12.7\,km\,s$^{-1}$ at the S$-$W edge of the H\,{\small II} region
labelled as 3b by \citet{hughes84} (Fig.1).  A single unresolved
maser component was detected in 22 spectral channel maps at the first
epoch, and in fewer channels 
at subsequent epochs.  The velocity, position (with 
relative errors arising from noise and fitting
uncertainties) and flux density of each maser component are given in Table 2 
at the three epochs.  All components brighter than 10$\sigma$
are included. At all three epochs 
the spectrum showed three features separated by
about 1.1\,km\,s$^{-1}$ from each other, with the strongest emission near
$-$15.4\,km\,s$^{-1}$ (Fig.1).  The emission comes  
from a region about 40\,mas in size, with three groups of maser 
components labelled A, B and C in Fig.1 that correspond to the three
spectral features. A is a single maser spot and B and C spots 
have angular sizes of the order of about 10\,mas.
Each group has a well defined central velocity.  
There is an overall velocity gradient along the arc$-$like structure 
from the north (blue-shifted emission) to the south west 
(red-shifted emission).  This structure was stable at all three epochs. 
At the first epoch the strongest 4765-MHz feature 
had a peak brightness of 1.55\,Jy\,beam$^{-1}$, corresponding to a lower 
limit on the brightness temperature of 
7$\times$10$^7$\,K.  The peak flux density $S_{4765}$ of all three features
showed a clear decrease over the time span of 8 weeks that can be 
modelled as 
\begin{equation}
S_{4765}\sim t^{\beta}
\end{equation}
where $t$ is the time in days since the start of the MERLIN observations. 
For all three features the mean value of $\beta$ was $-$0.34$\pm$0.08. 
 
We fitted a single Gaussian component to all spectral features, obtaining 
full width at half maximum intensity (FWHM) values ranging 
from 0.38\,km\,s$^{-1}$
to 0.82\,km\,s$^{-1}$. However, we did not find any relation
between the peak flux density and the FWHM. 

There was no difference between LHC and RHC polarization of the
4765-MHz maser emission within 10$\sigma$.  Data from the first epoch
were inspected for linearly polarized features and no emission was
found to a limit of about 8\,mJy. At the third epoch we detected
neither 4660-MHz nor 4750-MHz emission above  a level of 20\,mJy  
over the whole region searched. Therefore, the upper limits
for the integrated flux density ratios between three 6-cm OH lines
were $S(i)_{4660}:S(i)_{4765}\le$0.19 and
$S(i)_{4750}:S(i)_{4765}\le$0.18.

1720-MHz maser emission was found at both epochs of observation.
The parameters of components detected at the first epoch (above the
10$\sigma$ noise level) are given in Table 3. The emission appeared as
right and left completely circularly polarized features separated by
2.1\,km\,s$^{-1}$ (Fig.1). These components coincided within 23\,mas
in the sky so that there is no doubt that they are a Zeeman pair. The
central velocity of $-$14.34\,km\,s$^{-1}$ was close to the central
velocity of the 4765-MHz feature B. Assuming that the 1720-MHz
emission comes from well$-$separated $\sigma^{+1}$ and
$\sigma^{-1}$ components, with a line splitting of
118\,km\,s$^{-1}$G$^{-1}$, we estimate a magnetic field
strength of $-$17.3\,mG in the maser region. 
It is striking that the 1720-MHz emission coincided with group B of
the 4765-MHz maser components within 40\,mas.  Both lines were
detected at a position not previously noted for OH maser emission.
Moreover, within an area of size $2\arcsec\times2\arcsec$ centred at
the strongest component of group B we did not detect any emission
in the other three ground state OH lines above a sensitivity level of
15\,mJy.  The 1720-MHz component had a peak brightness of 
1.12\,Jy\,beam$^{-1}$, corresponding to a lower 
limit on the brightness temperature of 5$\times$10$^7$\,K.  1720-MHz
emission was also detected on 1999 June 16 within the same velocity
range and at the same positions, but we did not analyse these
data because the emission was too weak for self$-$calibration, being only 
half the intensity observed on 1999 May 20.    

\subsection{W75N}

Results for W75N are summarized in Figure 2 and Table 4. 
W75N showed a complex OH spectrum at 4765\,MHz with three
features within a velocity range of only 1.1\,km\,s$^{-1}$ (Fig.2). The
maser components found are listed in Table 4 where the velocities,
relative positions and the flux densities are given.  The 4765-MHz
components are located on the S$-$W edge of the H\,{\small II} region mapped 
in the radio continuum and labelled as Ba by \citet{hunter94} or VLA\,1 by
\citet{torrelles97} (Fig.2). 4765-MHz OH emission was detected from 
three regions A, B and C located in a chain of length
comparable to the synthesised beam size and roughly along the N$-$S direction
(Fig.2). There is a clear velocity gradient from the north (blue$-$shifted) 
to the south (red$-$shifted).  
The brightest component had a peak brightness of 
1.96\,Jy\,beam$^{-1}$, corresponding to a lower 
limit on the brightness temperature of 1.0$\times$10$^8$\,K.  
The position of the 4765-MHz maser structure
 coincides to within 60\,mas with the position of the 1720-MHz Zeeman pair 
Z$_{6}$ reported by \citet{hutawarakorn02}, while the demagnetized velocity of
the 1720-MHz Zeeman pair of 9.4\,km\,s$^{-1}$ is very close
(0.3\,km\,s$^{-1}$) to the velocity of feature A of the 4765-MHz
profile (Fig.2).  This implies that both maser transitions arise from
the same region and suggests similar pumping conditions.

\begin{table}
\caption{Components of OH 4765-MHz transition found in W75N. 
The J2000 coordinates of the brightest component are 
RA = 20$^{\rm h}$38$^{\rm m}$36\fs4238; Dec = 42\degr37\arcmin34\farcs484.}
\begin{tabular}{lrrcc}
\hline\hline
V$_c$        & $\Delta$RA & $\Delta$Dec & S$_{4765}$ & Group \\
(km\,s$^{-1}$) & (mas)        & (mas)         & (Jy\,b$^{-1}$) & \\
\hline
9.63  & 6.6(0.5)    &     27.1(0.5) & 0.270 & A \\
9.75  & 3.8(0.4)    &     22.0(0.5) & 0.316 & A \\
9.88  & 0.7(0.1)    &     2.2(0.1)  & 0.980 & B \\
10.00 & 0.0(0.1)    &     0.0(0.1)  & 1.957 & B \\
10.12 & 0.5(0.2)    &     0.2(0.2)  & 0.594 & B \\
10.25 & $-$5.8(1.3) & $-$14.3(1.3)  & 0.104 & C \\
10.37 & $-$2.5(0.2) & $-$12.7(0.2)  & 0.512 & C \\
10.49 & $-$2.3(0.2) & $-$13.0(0.2)  & 0.630 & C \\
10.62 & $-$2.1(0.5) & $-$14.8(0.5)  & 0.272 & C \\
\hline\hline
\end{tabular}                                
\end{table}

\begin{figure*}
\resizebox{\hsize}{!}{\includegraphics{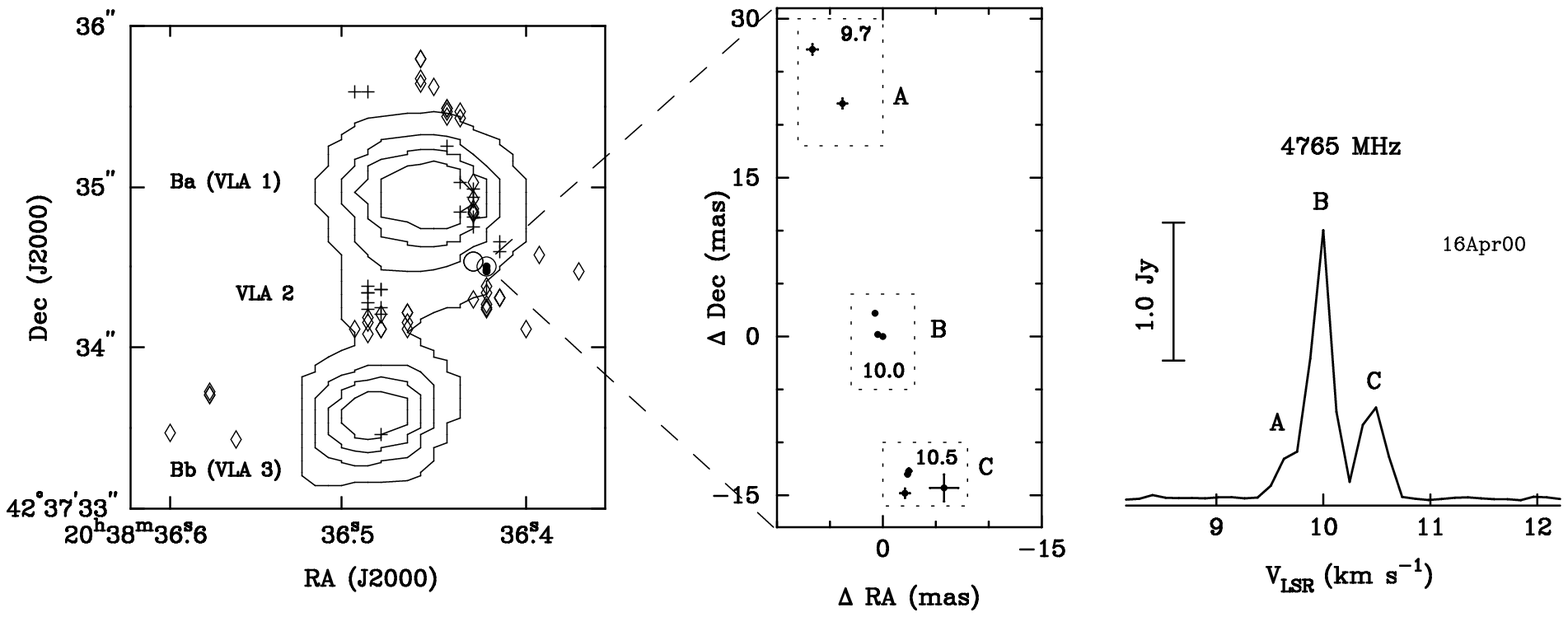}}
\caption{{\bf Left:} The map of the 8.4-GHz continuum emission of 
the W75N region \citep{hunter94}. The contours correspond 
to 10\%, 30\%, 50\% and 70\% of 2.5\,mJy\,beam$^{-1}$. 
The 4765-MHz masers are marked by dots.
Open circles and diamonds indicate the positions of the OH 1720-MHz
and 1665-MHz masers respectively \citep{hutawarakorn02},
while crosses mark  22-GHz  H$_2$O masers \citep{torrelles97}.
{\bf Middle:} Relative positions of OH 4765-MHz masers. The central velocity 
V$_{\rm LSR}$\,km\,s$^{-1}$ of each group and relative position errors 
of components are given. {\bf Right:} The total flux density spectrum 
of the OH 4765-MHz emission from W75N.}
\end{figure*}

\section{Discussion}
The important conclusion that can be inferred from the MERLIN observations 
of the two targets is that the 4765-MHz and 1720-MHz masers appear within  
the same velocity ranges and coincide in the sky well within the position 
uncertainties. 
This finding is especially valuable for Cep\,A as the observations of the 
two transitions were carried out at two epochs only 2.5 months apart.
For this source the position of group B of the 4765-MHz components
differs by less than 40\,mas from that of the 1720-MHz emission, while 
the central velocities of features at both transitions do not differ by 
more than 0.14\,km\,s$^{-1}$. It is therefore highly probable that 
the 4765- and 1720-MHz lines originate in the same gas volume. 

The flux density at 4765\,MHz decayed with time as given by
Equation 1 using $\beta=-0.34\pm0.08$ (Section 3.1). This gives an extrapolated 
flux density of 45\,mJy on 1999 May 20, the epoch of the 1720-MHz observation. 
This implies a ratio of $\sim$25 ($\pm$5) between the peak flux densities at
1720 and 4765\,MHz. 
Such an intensity ratio can be explained using the model by 
\citet{gray91} if there is no overlap between the pump line transitions 
for the different maser lines, assuming far infrared pumping.  
This occurs for regions with an H$_2$ number density, $n_{\rm H_2}$,  
of 4$\times$10$^6$\,cm$^{-3}$ and an OH number density, $n_{\rm OH}$,  
of 2$\times$10$^2$\,cm$^{-3}$ over a wide kinetic temperature range 
from 100 $-$ 200\,K. If there is a large-scale velocity field of up 
to 1.7\,km\,s$^{-1}$ across the masing region, line overlap allows 
the 1720-MHz maser to survive to higher densities with 
$n_{\rm H_2}$ up to a few $\times$10$^7$\,cm$^{-3}$. 

Under similar conditions saturated gain is predicted at both 1720- 
and 4765-MHz, as shown in Fig.12 of \citet{gray92}.  Their model also 
predicts the absence of the main line ground state OH masers and 
the other two 4.7-GHz lines.  This is fully consistent with our 
observations:  no emission was detected at 1665 or 1667\,MHz, 
with a typical rms of 5\,mJy in a projected area 
$2\arcsec\times2\arcsec$ centred at feature B of the 4765-MHz 
emission (Niezurawska et al., in prep.). Moreover, our upper limits 
to emission at 4660- and 4750\,MHz show that the 4765-MHz emission 
is at least 5 times stronger.  We note that the gain length of the 
4765-MHz emission calculated by \citet{gray92} is of the same order 
as the projected linear size of 4$\times$10$^{14}$\,cm of maser 
structure in Cep\,A for the assumed distance of 725\,pc.

Our lower limit for the brightness temperature of 4765-MHz
masers of 7$\times$10$^7$\,K in Cep\,A does not preclude saturation.  Very
Long Baseline Interferometry 
observations of W3(OH) and two other star forming regions have revealed
angular sizes of 4765-MHz components as small
as a few mas, implying brightness temperatures up to 10$^9$K
(\citealt{baudry88}; \citealt{baudry91}) which can be reached due to
saturated amplification. The relationship between the linewidth
and the peak flux density for feature A established from single$-$dish
observations \citep{szymczak00} appears to be consistent with a model
of saturated maser emission.

The newly detected 1720/4765-MHz OH maser components in Cep\,A are
located 2\farcs8 to the south$-$west of the previously known 
clusters of 1665-and 1667-MHz masers observed by \citet{cohen84}; 
\citet{migenes92} and
\citet{argon00}.  Comparison with the 1.3-cm continuum map reported by
\citet{torrelles98} reveals that our OH masers lie about 0\farcs5 to
the S$-$W of the centre of the  weak (2.3\,mJy) continuum source 
(labelled as 3b in Fig.1) elongated at a position angle of about 50\degr. 
They are accompanied by four weak ($<$0.3\,Jy) 22-GHz H$_2$O maser 
components grouped in a velocity range from $-$14.8\,km\,s$^{-1}$ to
$-$12.8\,km\,s$^{-1}$ and aligned almost parallel to the source axis.
Not obvious velocity gradient is seen along the water maser emission
\citep{torrelles98}.
The 1720/4765-MHz masers lie just at the S$-$W edge of this structure
0\farcs16 from the nearest $-$12.8\,km\,s$^{-1}$ water component.

The overall arc$-$like structure of the 4765-MHz maser in 
Cep\,A has a projected size of 4$\times$10$^{14}$\,cm and shows a 
velocity gradient along the axis of continuum source 3b, with a velocity 
dispersion of 3.1\,km\,s$^{-1}$. The adjacent 22-GHz water components 
exhibit a velocity dispersion of 2\,km\,s$^{-1}$ over a projected size of
4$\times$10$^{15}$\,cm.  These data clearly suggest a kinetically
complex medium where the velocity field may be dominated by turbulence
(e.g.\,\citealt{field82}).  The presence of turbulence in Cep\,A has
also been demonstrated by proper motion studies \citep{migenes92}.

The source labelled 3b in Fig.1 is characterized by a flat
spectral index between 1.5-GHz and 15-GHz indicative of
optically thin thermal emission \citep{garay96} which may arise from
shock heated gas excited by an external source. The adjacent sources
2, 3c and 3d show positive spectral indices $\alpha$
(convention $S\propto\nu^{\alpha}$) suggesting an internal source of
energy. It has been postulated that source 3b may be at the
earliest evolutionary stage among these four sources 
\citep{torrelles98}.  However, our detection of the 1720/4765-MHz
masers does not support that. The presence of weak H$_2$O
masers \citep{torrelles98} and OH masers in two lines 
has been interpreted as being due to object 3b being ionised by
an embedded B2 ZAMS star \citep{garay96}.
If source 3b harbours a star then for the assumed distance of 725\,pc
the excited OH masers arise at a projected separation of
5$\times$10$^{15}$\,cm. The edge of the molecular cloud, where
$n_{\rm H_2}\sim2\times10^{4}$\,cm$^{-3}$, hosts a magnetic field of strength
$B=0.3$\,mG, which scales  with number density as
$B\sim n^{0.4 - 0.5}$ (\citealt{garay96} and references therein),
consistent with compression of the interstellar magnetic field during
star formation.    
Our measurements of 1720-MHz Zeeman splitting (Section 3.1) imply 
a magnetic field strength of at least 17.3\,mG in the OH maser 
region.  This leads to an estimate of $n_{\rm H_2}\simeq7\times10^{7}$\,cm$^{-3}$. 
This value agrees well with the theoretical prediction for the 
1720-/4765-MHz masers \citep{gray92}. 

The 4765-MHz emission from Cep A showed pronounced variability, 
decreasing by a factor of 4 over a period of 7 weeks.  
The MERLIN data are consistent with single dish data previously  
reported by \citet{szymczak00}, assuming  that feature A corresponds 
to the peak discovered in the single$-$dish observations.  More
recent data taken with the Toru\'n antenna suggest the maser flux
density continues to decline to 0.2\,Jy or below (Fig.3). The
temporal characteristic of the 4765-MHz emission of Cep\,A seem to
resemble that observed in Mon\,R2 \citep{smits98}. 

It seems that the 1720-MHz maser also varies 
on a time scale of weeks.  It is remarkable
that Cep\,A showed no 1720-MHz emission for many years throughout the
1980's, with upper limits of 0.1 to 0.3\,Jy being regularly observed
(e.g.\,\citealt*{cohen90}).  The emission turned on some
time between July 1994, when the upper limit was 0.3\,Jy (Masheder \&
Cohen, in preparation), and June 1995, when a pair of emission features
appeared at $\sim$2\,Jy with velocities of $-$15.9 and
$-$13.9\,km\,s$^{-1}$ (Cohen, unpublished data).

\begin{figure}
\resizebox{\hsize}{!}{\includegraphics{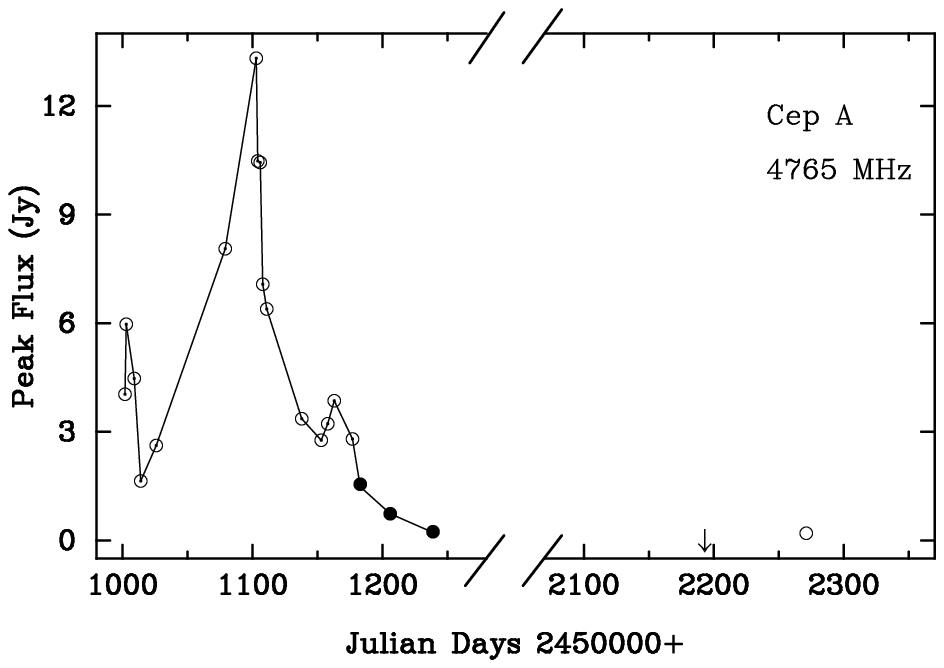}}
\hskip2.cm
{{\bf Fig. 3.} Variability of the strongest feature (A) of the 4765-MHz line 
in Cep A over the period from 1998 July 07 to 2001 December 27. The open
circles correspond to the single dish data 
\citep{szymczak00} and unpublished observations, the 
arrow marks an upper limit, and dots represent the MERLIN data analysed in 
this paper.}
\end{figure}

The 4765-MHz maser emission discovered towards W75N coincides
in velocity ($<$0.1\,km\,s$^{-1}$) and position ($<$60\,mas) with the
1720-MHz maser mapped with MERLIN about 6.5 years earlier by
\citet{hutawarakorn02}, the Zeeman pair Z$_6$.  The 4765-MHz
components in W75N are located between two clusters of 1665-MHz OH
masers at mean velocities of $V_{\rm LSR}$=7.3\,km\,s$^{-1}$
(northern) and $V_{\rm LSR}=$10.6\,km\,s$^{-1}$ (southern) labelled as
3 and 4 by \citet{baart86}. The 22-GHz H$_2$O masers at
10.7\,km\,s$^{-1}$ and 12.6\,km\,s$^{-1}$ reported by
\citet{torrelles97} lie about $\sim$200~mas north of the group B of
4765-MHz maser emission. The 1720-/4765-MHz components of W75N are
projected at the S$-$W edge of the H\,{\small II} region labelled Ba (VLA1) by
\citet{hunter94}, which has a spectral index of
about 0.7 between 8.5-GHz and 22-GHz.   This indicates partially
optically thick emission from an ionized biconical jet
\citep{torrelles97}.  The 1720-/4765-MHz masers lie at the S$-$W edge
of a molecular outflow at distance of 0\farcs54 from the centre of the
22-GHz continuum emission.  For the assumed distance of W75N of 2\,kpc
the projected distance of the 1720-/4765-MHz masers from a central star
is 1.6$\times$10$^{16}$cm.

The OH maser emission from W75N has been modelled in detail by 
\citet*{gray03}. Their model predicts 4765-MHz emission from +7 to
+12\,km\,s$^{-1}$ and 1720-MHz emission from +12 to +13\,km\,s$^{-1}$
(Figs. 4 and 5 in that paper), in good agreement with the
velocities and positions found here.  The physical conditions in 
the region where 1720-MHz and 4765-MHz coincide in the model are as
follows: $n_{\rm H_{2}}$ = 5$\times$10$^{8}$cm$^{-3}$, $n_{\rm OH}$ =
5$\times$10$^{3}$cm$^{-3}$, kinetic temperature $T_{\rm K}$ = 9.7~K, dust
temperature $T_{\rm d}$ = 45~K.  These are much denser and colder than the
conditions thought to prevail in the typical OH-H\,{\small II} region W3(OH)
where the amplification paths are much longer, the OH masers are much
more powerful and intense, and are thought to lie in a shocked
molecular envelope surrounding the expanding H\,{\small II} region.  In W75N
the molecular gas is cooler and denser and 
is associated with a rotating molecular disc and
bipolar outflow, at an earlier evolutionary phase than W3(OH).

In both sources we found the association of 1720- and 4765-MHz OH
masers that strongly supports the model by \citet{gray92}. In contrast, the
modelling of \citet{pavlakis96a}; \citet{pavlakis96b} and \citet{cragg02}
poorly fits our data.

A strong association between 4765- and 6035-MHz OH transitions has been
reported recently by \citet{dodson02}. The 6031- and 6035-MHz OH masers were
detected towards Cep\,A in a velocity range from $-$10 to $-$8\,km\,s$^{-1}$
\citep{baudry97}. Towards W75N they found only the 6035-MHz emission in a
velocity range from 7 to 8\,km\,s$^{-1}$. These data imply that in both
sourcs the 6031-/6035-MHz masers do not coincide in velocity with the
1720-/4765-MHz masers. No information is available where the 6031-/6035-MHz
masers are located.

\section{Conclusions}
Two star$-$forming regions Cep\,A and W75N have been mapped in the 4765- and
1720-MHz maser lines. Components of the excited OH emission at
4765\,MHz form arc and linear structures of projected size
4$\times$10$^{14} -$ 1$\times$10$^{15}$\,cm with a very well
defined velocity gradient. These structures lie at projected
distances of $0.5 - 2\times 10^{16}$\,cm from the central star and its
surrounding H\,{\small II} region.
In both sources the two maser lines show a good spatial coincidence
within the uncertainty of $\sim$35\,mas and demagnetised
velocity pairing within the 0.2\,km\,s$^{-1}$ spectral resolution,
implying that the 4765-/1720-MHz masers originate from the same
gas volume. No emission of any other 1.6- and 4.7-GHz OH transitions
was found in these regions.  Including the present data there
are now three examples of co-propagating 1720-MHz and 4765-MHz masers,
where the positional association has been verified to within
$\sim$30~milliarcseconds.

\vspace*{0.5cm}
\noindent
{\bf ACKNOWLEDGMENTS} \\
The authors thank Dr.\,M.\,Gray for discussions and 
Dr.\,G.\,Garay and Dr.\,T.\,Hunter for making available the
continuum maps of Cep A and W75N.
MERLIN is a national facility operated by the University 
of Manchester at Jodrell Bank on behalf of PPARC.
AN acknowledges a fellowship funded by the EU under the 
Marie Curie Training Site programme.
The work was supported by grant 2P03D01122
of the Polish State Committee for Scientific Research.


\label{lastpage}
 
\end{document}